\documentclass{aa}
\usepackage{graphicx}
\usepackage{amsmath}
\usepackage{amssymb,amsfonts}
\usepackage[mathscr]{eucal}
\usepackage{astron}
\usepackage{epsfig}

\begin{document}

   \thesaurus{} 

   \title{Correlation of the South Pole 94 data with $100\mu\mathrm{m}$ and 408 MHz maps}

   \subtitle{}

    \author{J.-Ch.~Hamilton\inst{1}, K.M. Ganga\inst{1,2}}

   \offprints{hamilton@cdf.in2p3.fr}
   \mail{}
   \institute{Physique Corpusculaire et Cosmologie,
              CNRS-IN2P3, 
              Coll{\`e}ge de France,
              11 pl. M. Berthelot,
              75231 Paris Cedex,
              France
              \and      
              Infrared Processing and Analysis Center,
              California Institute of Technology,
              Pasadena, CA \ 91125,
              USA}

   \date{December 29, 2000}

   \maketitle

   \begin{abstract}
     We present a correlation between the ACME/SP94 CMB anisotropy
     data at 25 to 45~GHz with the IRAS/DIRBE data and the Haslam
     408~MHz data.  We find a marginal correlation between the dust
     and the Q-band CMB data but none between the CMB data and the
     Haslam map.  While the amplitude of the correlation with the dust
     is larger than that expected from naive models of dust emission
     it does not dominate the sky emission.  
   \end{abstract}


\section{Introduction}

The study of Cosmic Microwave Background (CMB) anisotropies has
recently proven to be a powerful tool for observational cosmology
\cite{boomerangomega,boomerang,maxima,maxima_param}. One of the most important
aspects of CMB analyses is to check whether the anisotropies observed
are due to CMB temperature fluctuations or to foreground contamination
such as diffuse Galactic emission.

Diffuse Galactic emission is dominated at high frequency (above
$\approx 100$~GHz) by thermal emission from dust. High quality tracers
of this emission are given by the IRAS/DIRBE $100~\mu\mathrm{m}$ maps
and have been made available in a user friendly way by \cite{SFD}. At
lower frequencies, Galactic emission is dominated by synchrotron and
free-free radiation. Our best tracer of the synchrotron emission of
the Galaxy is given by the Haslam 408 MHz map \cite{haslam}. Free-free
emission is traced by H$\alpha$ emission, but maps of this emission
are not yet publicly available.

\begin{figure}
\resizebox{\hsize}{!}{\epsfig{file=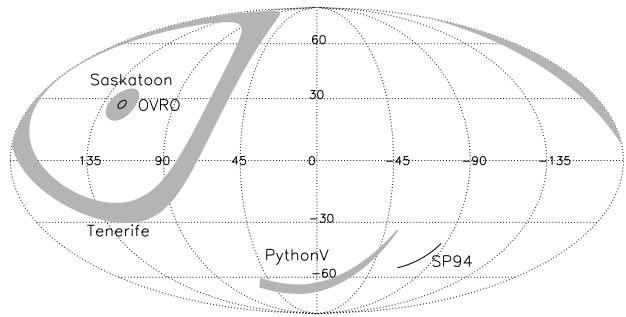}}
\caption{Regions of the sky (in Galactic coordinates) covered by the
  different experiments that have been correlated with dust. Saskatoon
  is the grey region around the North Celestial Pole. OVRO observed 36
  small patches of the sky at 2 degrees from the Pole. The main part
  of the Python~V coverage is the grey region in the center of the
  lower part of the map. Tenerife observed a 10 degrees width strip
  around the North Celestial Pole (grey ring).  COBE and 19~GHz both
  covered the whole sky. South Pole is at the right of Python V.
\label{regions}}

\end{figure}

Characterizing the correlations between these different Galactic
components and data taken in the microwave is important in order to
understand the spectral behaviour and the origin of Galactic emission
which may contaminate CMB measurements. The first significant,
high-$|b|$ cross-correlation between the COBE/DMR maps and dust
templates was found by \cite{kogut96,kogut96b}. Significant
correlations were found at each DMR frequency, but with a spectral
behaviour in better agreement with free-free emission than with
vibrational dust. These results have been confirmed by different
experiments in various parts of the sky:
Saskatoon~\cite{doc_saskatoon}, OVRO~\cite{leich97} and
19~GHz~\cite{doc19ghz}.  At high Galactic latitudes, however,
Python~V~\cite{pythonV} did not see any correlation.  The regions of
the sky covered by these experiments are shown in Fig.~\ref{regions}.
As noted above, free-free emission should correlate with H$\alpha$
maps. Unfortunately, correlations found between H$\alpha$ maps and the
CMB indicate that the H$\alpha$-traced emission is too small to
explain all of the observed correlation between CMB data and
$100~\mu\mathrm{m}$ data.

\cite{draine_laz98a,draine_laz98b} have suggested an alternate
explanation for the correlation between CMB data and dust maps, namely
that this emission may be the result of rotational dust emission from
elongated grains.  This emission could be compatible with the observed
spectrum of the dust correlated emission in the microwave frequencies.

The sum of dust components, both rotational and vibrational, should
show a local minimum at roughly 70~GHz, increase to a peak around
10~GHz, and drop off at lower frequencies. Such behaviour has been
observed by \cite{doc_spindust} in correlating the Tenerife 10 and 15
GHz data with the IRAS/DIRBE maps. This has been put forth as evidence
supporting spinning dust as the origin of the dust-correlated
emission. A recent analysis of yet more Tenerife data
\cite{mukherjee00} reports, however, that this result should be
considered with care, as it originates mainly from a small, high
Galactic latitude region. They finally say that the data does not
allow any conclusion concerning the origin of the dust correlated
emission.

In this article we present a similar analysis using the South Pole
1994 (SP94) data~\cite{gundersen95}. As correlations in the North
Celestial Pole region have recently been questioned, we would like to
obtain another measure of the correlation in a different region of the
sky. More precisely, as correlations were found at low Galactic
latitude (North Celestial Pole) and no correlation was found at high
Galactic latitude by Python~V, we wanted to explore the correlations
in the high latitude regions that will be important for MAP and
Planck.  SP94 was a CMB experiment that used the Advanced Cosmic
Microwave Explorer (ACME) which looked at such a high-$|b|$ region.

\section{Data}
\subsection{Microwave data}
The SP94 data were taken in two different bands with 7 different
channels: the K$_a$-band with four center frequencies (27.25, 29.75,
32.25 and 34.75 GHz) and the Q-band with three different frequencies
(39.15, 41.45 and 43.75 GHz). A detailed description of the instrument
can be found in \cite{meinhold93}, while the detail of the 1994
observations and data reduction is described in \cite{gundersen95}.
Frequencies and beam width for these channels are given in
Table~\ref{tab_beams}.
\begin{table}
\begin{center}
\begin{tabular}{ccc}\hline
Channel &Frequency (GHz) &Beam FWHM (deg.)\\
\hline 
K$_a$1 &27.25 &1.67\\
K$_a$2 &29.75 &1.53\\
K$_a$3 &32.25 &1.41\\
K$_a$4 &34.75 &1.31\\
Q1 &39.15 &1.17\\
Q2 &41.45 &1.11\\
Q3 &43.75 &1.05\\
\hline 
\end{tabular}
\end{center}
\caption{Frequencies and beam Full-Width-at-Half-Maximum for the
different channels.}
\label{tab_beams}
\end{table}

\subsection{Infrared data}
In this analysis, we used two templates for each channel :
\begin{itemize}
\item For the thermal emission from dust, we used the IRAS/DIRBE map
  provided by \cite{SFD}. The initial resolution of this map is 6.1
  arcmin FWHM.
\item For the synchrotron emission, we used the Haslam maps
  \cite{haslam} whose resolution is 51 arcmin FWHM.
\end{itemize}

\begin{figure}
\resizebox{\hsize}{!}{\epsfig{file=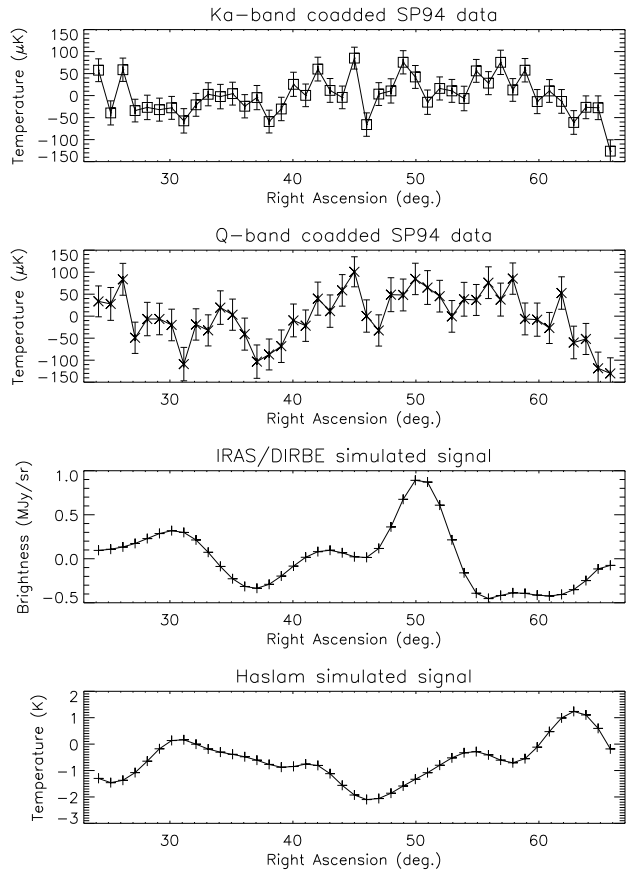}}
\caption{The top panels shows the coadded SP94 data for the Q-band and
  for the K$_a$-band separately. The coaddition was performed using,
  for each pixel, the full covariance matrix (using the method
  explained at the end of section \ref{sect_method}). The third panel
  shows the simulated IRAS/DIRBE data (dust) and the bottom panel
  shows the simulated Haslam data (synchrotron). For both simulated
  signals, only the 27.25~GHz band is shown, the others are very
  similar. \label{data_simu}}
\end{figure}

Both maps were convolved to the SP94 beams (for each channel
separately) using the beams given in Table~\ref{tab_beams} and then
differenced, simulating the chopping strategy of the SP94
experiment. The simulated signals are shown, along with the SP94 data
in Fig.~\ref{data_simu}. We see from this figure that the
correlations, if non-zero, will be rather small.

\section{Method \label{sect_method}}
For each of the seven frequency bands we have $N=43$ data points,
all at a declination of $-62^\circ$ and ranging from $23^\circ$ and
$67^\circ$ in right ascension (see Fig.~\ref{regions}).
These data points were obtained using a sinusoidal $1.5^\circ$ chop
with smooth, constant declination, constant velocity scans. We
directly follow in this section the notation of
\cite{doc_spindust}. We assume that the data is a linear combination
of CMB anisotropies $x^i_{CMB}$ and $M$ foreground components
(including an unknown offset and gradient) that are given by simulated
observations of the template dust and synchrotron maps using the SP94
beam and scanning strategy. In a vector-like notation, one gets :
\begin{equation}
{\bf y}=X\vec{a}+\vec{x}_{CMB}+\vec{n}
\end{equation}
$X$ is a matrix with $N$ rows and $M$ columns. It contains the
template maps observed using the SP94 beams and scanning strategy.
$\vec{a}$ it the $M$-vector we are interested in, it contains the
correlation of the SP94 data with the different foregrounds contained
in $X$. The experimental noise is $\vec{n}$.

The noise and the CMB anisotropies are each assumed to be uncorrelated
Gaussian variables with zero mean. They each have non-trivial
covariance matrices and together a total covariance matrix of:
\begin{eqnarray}
C&=&\left<\vec{y}\vec{y}^T\right>-\left<\vec{y}\right>\left<\vec{y}^T\right>
\\
&=&
\left<\vec{x}_{CMB}\vec{x}_{CMB}^T\right>+\left<\vec{n}\vec{n}^T\right>
= C_{CMB} + C_n
\end{eqnarray}

As discussed in~\cite{gundersen95}, the noise in a given pixel in
given band (K$_a$ or Q) is only correlated with the noise in the same
pixel in the different channels of this band (4 channels for the K$_a$
band and 3 channels for the Q band). The noise from the K$_a$ and Q band
are uncorrelated.

The covariance matrix of the CMB is obtained through the window
function of the South Pole experiment $W_{\ell,ij}^{kl}$ that can be
found in \cite{gundersen95} :
\begin{equation}
C_{CMB}~_{ij}^{kl}=\sum_{\ell=0}^\infty\frac{2\ell+1}{4\pi}C_\ell
W_{\ell,ij}^{kl}
\end{equation}
where $\ell$ is the multipole index, $i$ and $j$ denote pixel indices
and $k$ and $l$ denote the different channels.  The window function
was calculated accounting for the different beam-widths for each
channel, and the underlying CMB sky was assumed to be the same for all
channels.  The beam width of each channel can be found in Table
\ref{tab_beams}.  $C_l$ is the CMB anisotropy angular power spectrum.
We used the best fit $\Omega=1$ model from \cite{jaffe:2000} obtained
with the combined COBE/DMR, BOOMERANG and MAXIMA-1 data sets:
$\Omega_\mathrm{tot}=1$, $\Omega_\Lambda=0.7$, $\Omega_bh^2=0.03$,
$\Omega_ch^2=0.17$, $n_s=0.975$, $\tau_C=0$. As a check, we confirmed
that the results do not change much using a flat band power spectrum
with $Q_\mathrm{rms-PS}=20$, $25$ and $30~\mu\mathrm{K}$. The
differences were small.

We are interested in measuring the correlation coefficients with the
various templates, we therefore want to measure $\vec{a}$ considering
$\vec{n}$ and $\vec{x}_{CMB}$ as noise (but accounting for chance
alignment between CMB and the templates trough the CMB covariance
matrix). We therefore construct the following $\chi^2$ :
\begin{equation}
\chi^2=\left({\bf y}-X\vec{a}\right)^T C^{-1} \left({\bf
y}-X\vec{a}\right)
\end{equation}
Minimising this $\chi^2$ leads to the best estimate of $\vec{a}$ :
\begin{equation}
\hat{\vec{a}}=\left[X^T C^{-1} X\right]^{-1} X^T C^{-1} \vec{y}
\end{equation}
We make the fit simultaneously for all channels in order to account
for the channel-to-channel correlation matrix elements but the
templates (dust and 408~MHz emission) are fitted separately.  The
vector $\vec{y}$ contains the concatenated data for all channels ($N$
data points per channel).  It is therefore a $(n_{chan}\times N)$
vector.  The coadded Q-band and coadded K$_a$-band data are shown in
Fig. \ref{data_simu}. The covariance matrix is an $(n_{chan}\times
N)\times(n_{chan}\times N)$ matrix.  The $X$ matrix contains the
template simulated signal (one for each channel) as well as a constant
term and a gradient term for each channel, in order to account for the
fact that such terms were removed from the SP94 data. The matrix $X$
is therefore a $(n_{chan}\times N)\times(n_{chan}\times 3)$ matrix. In
a given column (corresponding to a given channel), $X$ contains mostly
zeros, except for the rows corresponding to that channel where it
contains the template simulated signal.

The best estimate of $\vec{a}$ is given through the $\chi^2$
minimization along with its covariance matrix:
\begin{equation}
\Sigma = \left[ X^T C^{-1} X \right]^{-1}\label{err_a}
\end{equation}
We obtain a correlation coefficient for each channel but those are
highly correlated according to the covariance matrix $\Sigma$. The
correlation coefficients cannot therefore be averaged in a
straightforward manner. We estimate the average of the correlation
coefficients via:
\begin{equation}
\chi^2=\left(\vec{a}-\bar{a}\vec{1}\right)^T \Sigma^{-1}
\left(\vec{a}-\bar{a}\vec{1}\right)\label{chisqa}
\end{equation}
where $\vec{1}$ is a vector whose elements are all equal to 1. The
best estimate of $\bar{a}$ is then obtained by:
\begin{equation}
\bar{a}=\frac{\mathrm{Total}(\Sigma^{-1}\vec{a})}{\mathrm{Total}(\Sigma^{-1})}\label{aval}
\end{equation}
where the $\mathrm{Total}$ function returns the sum of all elements of
a matrix or vector. This is equivalent to a weighted average, though
in this case we also account for correlations between the different
measures. The error-bar on $\bar{a}$ is then:
\begin{equation}
\sigma_{\bar{a}}=\sqrt{\left(\mathrm{Total}(\Sigma^{-1})\right)^{-1}}. \label{aerr}
\end{equation}
We do this for the K$_a$- and Q-bands separately, and for the
combination of the two. The results are shown in Table~\ref{tabcorrel}. 

Equations~\ref{chisqa},\ref{aval} and \ref{aerr} implicitly suppose
that the dust-correlated component has a flat spectrum ({\it ie} has a
spectral index $n=0$). As was mentioned before, results from other
experiments tend to favor spectral indices close to $n\simeq-2$
similar to free-free emission. In order to account for a non-zero
spectral index, we model the emission as :
\begin{equation}
  \bar{a}=\bar{a}_0\left(\frac{\nu}{\nu_0}\right)^n
\end{equation}
where $\nu_0$ is a reference frequency and not an additional degree
of freedom. Equation \ref{chisqa} then becomes :
\begin{equation}
  \chi^2=\left(\vec{a}-\bar{a}_0\vec{m}\right)^T \Sigma^{-1}
  \left(\vec{a}-\bar{a}_0\vec{m}\right)    
\end{equation}
where $\vec{m}$ is a vector with 7 elements
$m_i=\left(\frac{\nu_i}{\nu_0}\right)^n$.  This $\chi^2$ is not linear
if one lets $n$ vary as a free parameter. We therefore minimize it
numerically for a grid of 300 values of $n$ between -5 and 5.

\section{Correlations}
\begin{figure}
  \resizebox{\hsize}{!}{\epsfig{file=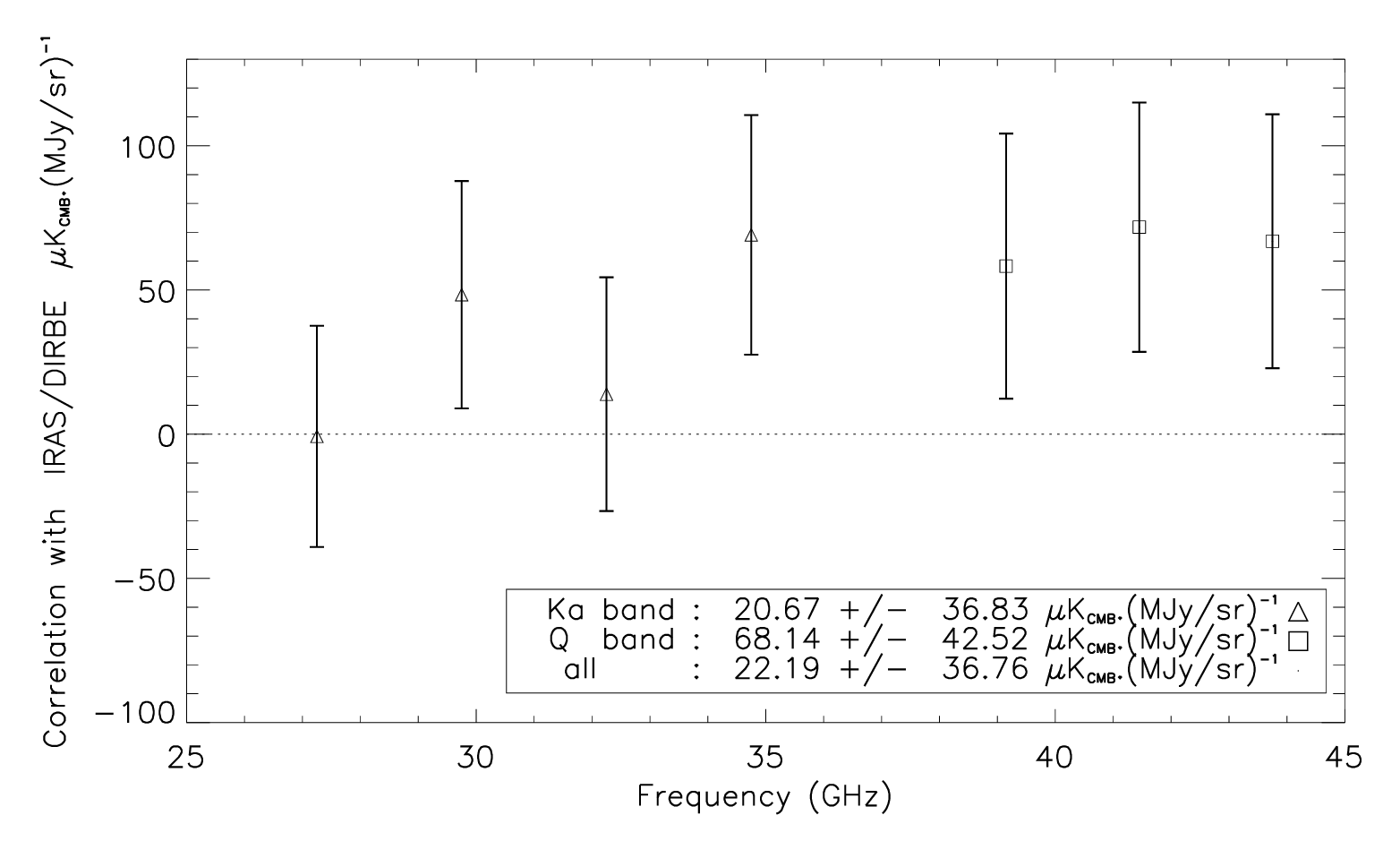}}
  \resizebox{\hsize}{!}{\epsfig{file=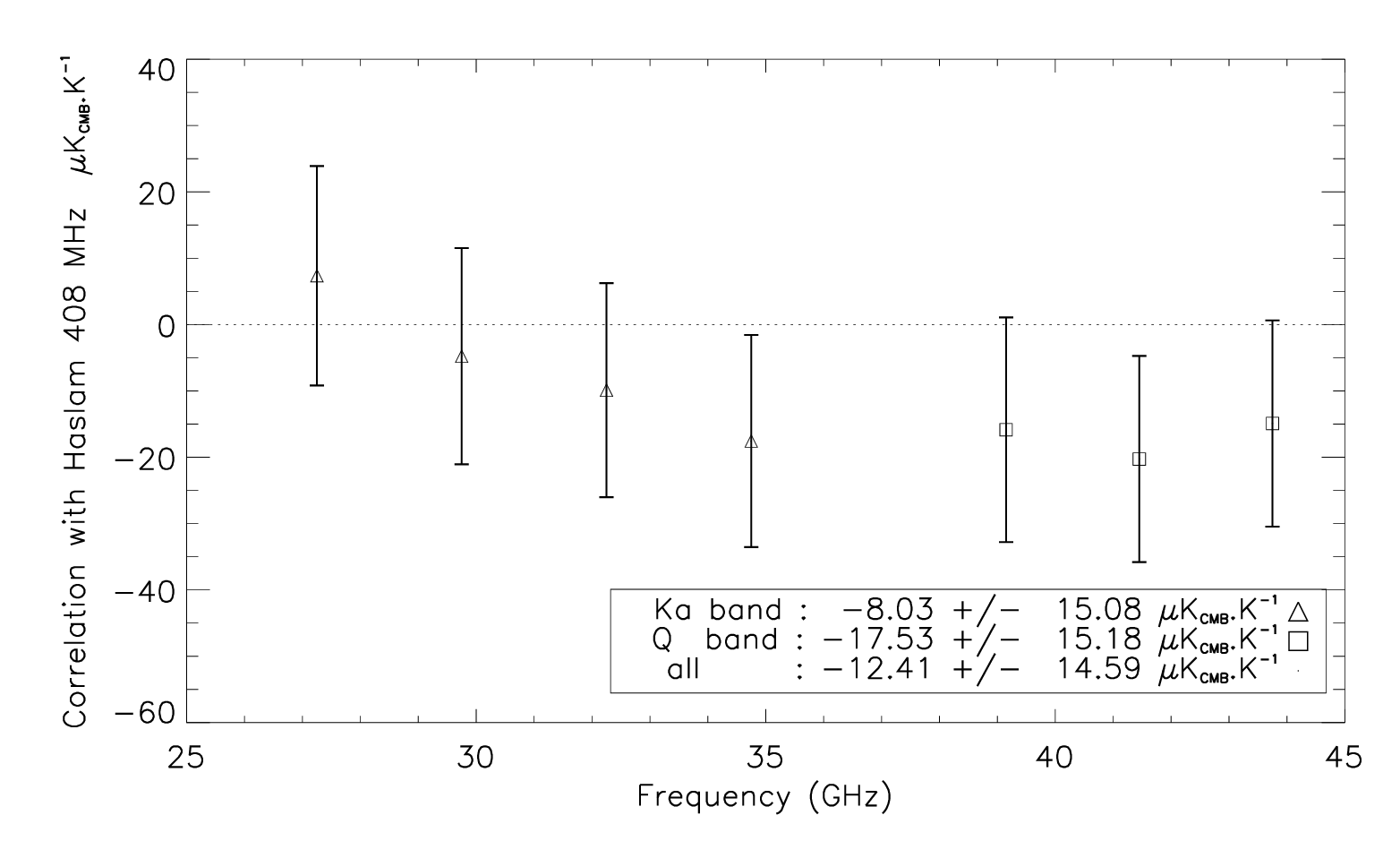}}
\caption{Correlations coefficients between the SP94 data and the 
  templates (top: with IRAS/DIRBE and bottom: with Haslam 408 MHz).
  The combined correlation coefficient assume a spectral index $n=0$
  for the correlated component.
\label{figres}}
\end{figure}

We compute correlations between the SP94 and both dust and synchrotron
templates shown in Fig.~\ref{data_simu} and Table~\ref{tabcorrel}.
Those results are also plotted as a function of the frequency in
Fig.~\ref{figres}.  The results we obtain show a marginal correlation
of the Q-band data with the dust IRAS/DIRBE template whereas there is
no correlation in the K$_a$-band. No significant positive correlation
is found in either band with the Haslam template.  The uncertainty on
the combined Q-band and K$_a$-band correlation coefficients is not as
much smaller than that of the Q-band or K$_a$-band alone as one might
naively expect. In fact, this is due to the contribution of the CMB
covariance matrix. For each channel, the CMB covariance matrix is
comparable to the noise one and therefore combining the data reduces
the noise contribution but not the CMB one that becomes dominant.

In Table~\ref{tabcorrel}, we also show the result for the analysis
allowing the spectral index to vary as a free parameter.  Combining
the K$_a$ channels did not lead to any realistic value for the
spectral index (the best fit, $n=5.0$, was at the edge of the range we
searched). Combining the Q-band data leads to a best fit for the
spectral index of $n=0.8\pm 2.0$. Combining all K$_a$ and Q data
together leads to $n=2.1\pm 1.0$. It is there clear that positive
values are preferred. This is not a surprise, however, as the Q-band
data obviously correlate with the IRAS data better than the K$_a$
does.

\begin{figure}
\resizebox{\hsize}{!}{\epsfig{file=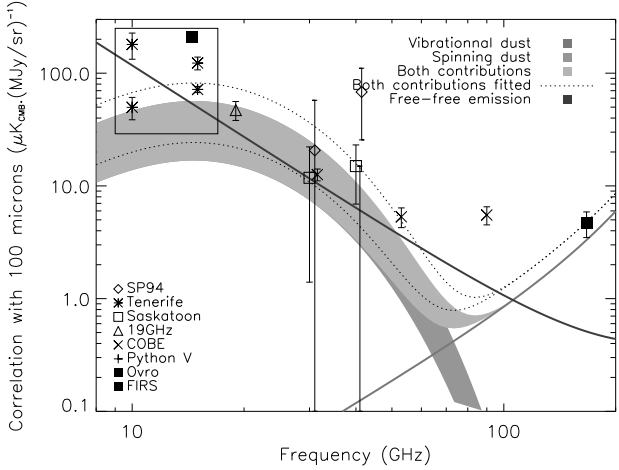}}
\caption{ CMB/IRAS correlation coefficients as measured by various 
  experiments. We overplot the spectra for vibrationnal dust, spinning
  dust and the sum of the two. The vibrationnal dust spectrum assumes
  a spectral index of 2.0 and was normalized to match averages of
  DIRBE at Galactic latitudes higher than 20 degrees. The spinning
  contribution is normalized following (Draine and Lazarian, 1998a).
  To guide the eye, we also plot the best fit of the models to the dta
  in dotted lines. We also added a free-free emission spectrum with a
  spectral index of -2.1 and an arbitrary normalization.
\label{resume_mes}}
\end{figure}

We summarize in Fig.~\ref{resume_mes} the measurements performed up to
now of the correlation with the vibrational dust traced by IRAS/DIRBE
$100\mu\mathrm{m}$ : SP94 corresponds to the present article, the
Tenerife measurements have been published in two different articles
\cite{doc_spindust,mukherjee00} reporting slightly different results
obtained with almost the same data. We plot both measurements, the
lower values were obtained by \cite{doc_spindust} and considered as a
possible evidence for the presence of spinning dust because of the
fall at 10 GHz. The upper measurements \cite{mukherjee00} were
published more recently and mitigated the enthusiasm for spinning
dust. As the values in this region are under question now, we plotted
a square around these points. OVRO measured one point in this region
at 14.5 GHz \cite{leich97} that seems to favour upper values of the
correlation.  Saskatoon data are taken from \cite{doc_saskatoon} and
show a $1\sigma$ detection in the K$_a$-band (not significant) and a
$1.9\sigma$ detection in the Q-band.  The 19~GHz (whole sky survey)
data were taken from \cite{doc19ghz}.  The result from PythonV at 41
GHz was taken from \cite{pythonV} and is fully compatible with no
detection (the best fit is negative and does not appear in the plot).
The COBE data points are from \cite{kogut96b}. The results quoted in
the article were obtained by fitting DIRBE $140\mu\mathrm{m}$ to the
DMR data. We therefore corrected them by the average ratio DIRBE
$140\mu\mathrm{m}$/DIRBE $100\mu\mathrm{m}$ to have them in the same
units as the other data points. The FIRS point at 167 GHz was obtained
by \cite{kmg_thesis}.  We also overplotted in Fig.~\ref{resume_mes}
the predicted spectrum (in terms of ratio to IRAS/DIRBE
$100\mu\mathrm{m}$) for vibrational dust in green (we assumed a
spectral index of $2.0$) normalised with IRAS/DIRBE
$100\mu\mathrm{m}$, and for spinning dust in blue. We also added the
spectrum for free-free emission in light blue (arbitrary normalization
and spectral index -2.1).  For the spinning dust, we follow
\cite{draine_laz98a} for the mixing between Warm Ionised Medium (WIM),
Warm Neutral Medium (WNM) and Cold Neutral Medium (CNM) models using a
respective fraction of 0.14, 0.43 and 0.43. The respective
normalization of these models was also taken from
\cite{draine_laz98a}. The sum of both contributions is shown in
Fig.~\ref{resume_mes} in red. The width of the curves for the spinning
dust model is due to the galactic latitude dependance of the optical
depth of the spinning dust components. We used all the latitudes more
than $20^\circ$ from the Galactic equator.  The normalization is that
given by \cite{draine_laz98a}. As clearly yields too little emission
to match the data, we have also done a fit to the points and
arbitrarily raised the model curves by this amount.  This is indicated
by the dotted curve.

\begin{table}
\begin{center}
\begin{tabular}{lrrcrc}\hline 
Band &$n$ &100 $\mu$m &sig. &408 MHz &sig.\\
\hline \hline
K$_a$1 &- &$-0.8 \pm 38.4$ &$-0.0$    & $7.4   \pm 16.5$ &$0.4$ \\
K$_a$2 &- &$48.4 \pm 39.4$ &$1.2$     & $-4.8   \pm 16.3$ &$-0.3$ \\
K$_a$3 &- &$13.9  \pm 40.5$ &$0.3$     & $-9.9  \pm 16.1$ &$-0.6$ \\
K$_a$4 &- &$69.1 \pm 41.5$ &$1.7$     & $-17.6  \pm 16.0$ &$-1.1$ \\
\hline
K$_a$  &-2.0 &$-16.6 \pm 20.2$ &$-0.8$     &- &- \\
K$_a$  &0.0  &$20.7 \pm 36.8$  &$0.6$      & $-8.0  \pm 15.1$ &$-0.5$ \\
K$_a$  &0.8  &$49.2 \pm 40.9$  &$1.2$      &- &- \\
K$_a$  &2.0  &$78.0 \pm 50.0$  &$1.9$      &- &- \\
K$_a$  &2.1  &$79.4 \pm 40.7$  &$2.0$      &- &- \\
\hline \hline
Q1  &- &$58.3 \pm 46.0$ &$1.3$     & $-15.9  \pm 16.9$ &$-0.9$ \\
Q2  &- &$71.8 \pm 43.2$ &$1.7$     & $-20.3 \pm 15.5$ &$-1.3$ \\
Q3  &- &$66.9 \pm 44.0$ &$1.5$     & $-14.9  \pm 15.5$ &$-1.0$\\
\hline
Q   &-2.0 &$73.5 \pm 54.2$ &$1.4$     & - &- \\
Q   &0.0  &$68.1 \pm 42.5$ &$1.6$     & $-17.5  \pm 15.2$ &$-1.2$ \\  
{\bf Q}   &{\bf 0.8}  &{\bf 60.1} $\pm$ {\bf 37.0} &{\bf 1.6}     & - &- \\
Q   &2.0  &$44.4 \pm 28.2$ &$1.6$     & - &- \\
Q   &2.1  &$42.8 \pm 27.3$ &$1.6$     & - &- \\
\hline 
\hline
All &-2.0 &$-25.3 \pm 20.2$ &$-1.4$     & - &- \\
All &0.0  &$22.2 \pm 36.8$  &$0.6$      & $-12.4  \pm 14.6$ &$-0.9$ \\
All &0.8  &$63.5 \pm 36.7$  &$1.7$     & - &- \\
All &2.0  &$56.8 \pm 24.8$  &$2.3$     & - &- \\
{\bf All} &{\bf 2.1}  &{\bf 54.2} $\pm$ {\bf 23.6}  &{\bf 2.3}     & - &- \\
\hline
\end{tabular}

\end{center}
\caption{Correlations coefficients between the SP94 data and the templates :  We also give the coadded correlation coefficients found 
  by combining the K$_a$ channels, the Q band channels and for all channels 
  together. The second column gives the assumed correlated signal spectral 
  index in each case (we considered $n\neq 0$ only for the correlation with 
  dust). The third  column shows the coefficient of the correlation with
  the dust IRAS/DIRBE  simulated signal (in 
  $\mu\mathrm{K}.(\mathrm{MJy/sr})^{-1}$) for each spectral index and the 
  fifth column shows this
  correlation for the Haslam 408 MHz simulated signal (in 
  $\mu\mathrm{K}.K^{-1}$). The fourth and sixth columns indicate the 
  significance in terms of number of sigma of the correlation measured. 
  No best fit was found in the range $-5 \le n \le 5$ for the combined K$_a$
  channels. For the combined $Q$ channels, the best fit was found for $n=0.8$
  and is shown in bold. Combining all the channels leads to a best fit at 
  $n=2.1$ also shown in bold.}
\label{tabcorrel}
\end{table}

\section{Discussion and conclusion}
The signal to noise for the Q-band detection is only 1.6 ($89\%$
confidence level) so this would not be considered a detection as most
CMB experiments require at least 2$\sigma$. We would note however
that, as opposed to normal CMB anisotropy analyses, here we are
comparing the data to a template and it is difficult to imagine that
systematic effects or analysis errors that would cause a random
correlation with $100\mu\mathrm{m}$ dust emission. The correlation
could arise from random alignment between the CMB anisotropies and the
dust template. This, however, should be taken into account in our
analysis if our covariance matrices for the data and the CMB are
correctly estimated. If these covariance matrices were underestimated,
the error bars we compute on the correlation coefficient (with
Eq.~\ref{err_a}) would be too small and therefore the significance of
our result would be overestimated.  In order to check the validity of
our error bars, we correlated the SP94 data with template dust maps
obtained by rotating the initial template maps around the Galactic
poles and by inverting North and South. The Galaxy was either rotated
and/or inverted in 36 different ways (10 degrees each) to make 36
different simulations. We found a zero average correlation and the RMS
of the correlations normalized by the error bars (calculated with
Eq.~\ref{err_a}) appears to be 1.1. This shows that the covariance
matrices of the CMB and the data and therefore our error bars are
correctly estimated, confirming the significance of the correlation
coefficients we obtain.

We do, however, find the following points interesting: As seen
previously \cite{doc_saskatoon} the correlation between the Q-band and
the $100\mu\mathrm{m}$ emission is stronger than the correlation
between the K$_a$-band and the $100\mu\mathrm{m}$ emission. If the
correlation we found is to be believed, the ratio of the RMS of the
$100\mu\mathrm{m}$ template times the fitted correlation coefficient
to the implied sky RMS is $0.38$ (in the Q-band).  This result
indicates that roughly $14\%$ of the power seen on the sky by
ACME/SP94 Q-band could be due to Galactic emission. The $C_\ell$ could
go down by $14\%$ and the amplitude by $38\%$. This however does not
apply to the K$_a$-band. This result is in qualitative agreement with
\cite{gundersen95} and \cite{kmg_acme}, both of which found different
spectral indices for the K$_a$- and Q-band data, though again, with
low statistical significance.

We have also done the above analysis using as a template not the raw
$100\mu\mathrm{m}$ data but rather the extrapolations recommended by
\cite{SFD} (model 8) to 500GHz and 40 GHz. The $\chi^2$ for the fit
went down by a very small amount. The differences in the $\chi^2$ were
not significant and do not change the conclusions drawn above. This is
not surprising as the SP94 data covers only a small region and the SFD
model was designed to cope with the whole sky.

In conclusion, we have shown that there is a very marginal correlation
between the SP94 Q-band data and vibrational dust tracers, while no
such correlation exists with the 408 MHz map. The amplitude of the
correlation in the Q-band is larger than that seen by
\cite{doc_saskatoon}. 
                       
\begin{acknowledgements}  
  The authors would like to thank J.O.~Gundersen and the ACME team for
  making the SP94 data available. We would also like to thank 
  F.X.~D{\'e}sert for fruitful discussions.

\end{acknowledgements}

\end{document}